\documentclass[a4paper,11pt]{article}
\usepackage{pos}
\usepackage[english]{babel}           
\usepackage{color}                    
\usepackage{titlesec}                 
\usepackage{newlfont}          	      
\usepackage[T1]{fontenc}              

\usepackage{tikz}                     
\usepackage{tikz-feynman}             
\tikzfeynmanset{compat=1.1.0}         
\usetikzlibrary{bending}              
\usetikzlibrary{decorations.markings} 
\usetikzlibrary {arrows.meta}         
\usetikzlibrary{snakes}               
\usepackage{graphicx}                 
\usepackage{pdfpages}                 

\usepackage{amsmath}                  
\usepackage{dsfont}                   
\usepackage{simpler-wick}             

\pgfdeclaredecoration{Snake}{initial}
{\state{initial}[switch if less than=+.625\pgfdecorationsegmentlength to final,
                  width=+.3125\pgfdecorationsegmentlength,
                  next state=up]{
    \pgfpathmoveto{\pgfqpoint{0pt}{\pgfdecorationsegmentamplitude}}
  }
  \state{down}[switch if less than=+.8125\pgfdecorationsegmentlength to end down,
               width=+.5\pgfdecorationsegmentlength,
               next state=up]{
    \pgfpathcosine{\pgfqpoint{.25\pgfdecorationsegmentlength}{-1\pgfdecorationsegmentamplitude}}
    \pgfpathsine{\pgfqpoint{.25\pgfdecorationsegmentlength}{-1\pgfdecorationsegmentamplitude}}
  }
  \state{up}[switch if less than=+.8125\pgfdecorationsegmentlength to end up,
             width=+.5\pgfdecorationsegmentlength,
             next state=down]{
    \pgfpathcosine{\pgfqpoint{.25\pgfdecorationsegmentlength}{\pgfdecorationsegmentamplitude}}
    \pgfpathsine{\pgfqpoint{.25\pgfdecorationsegmentlength}{\pgfdecorationsegmentamplitude}}
  }
  \state{end down}[width=+.3125\pgfdecorationsegmentlength,
                   next state=final]{
     \pgfpathcosine{\pgfqpoint{.25\pgfdecorationsegmentlength}{-1\pgfdecorationsegmentamplitude}}
     \pgfpathsine{\pgfqpoint{.25\pgfdecorationsegmentlength}{-1\pgfdecorationsegmentamplitude}}
  }
  \state{end up}[width=+.3125\pgfdecorationsegmentlength,
                 next state=final]{
     \pgfpathcosine{\pgfqpoint{.25\pgfdecorationsegmentlength}{1\pgfdecorationsegmentamplitude}}
     \pgfpathsine{\pgfqpoint{.25\pgfdecorationsegmentlength}{1\pgfdecorationsegmentamplitude}}
  }
  \state{final}{\pgfpointdecoratedpathlast}
}

\newlength{\wdth}
\newcommand{\self}[2]{\settowidth{\wdth}{#2}#2\hspace{-.5\wdth}\c#1{\vphantom{#2}}\c#1{\vphantom{#2}}\hspace{.5\wdth}}

\newdimen\wickgap \wickgap=2pt 
\newbox\wickbox
\def\doublewick\c#1#2\c#3{\setbox\wickbox=\hbox{$#2$}
 \c#1{\phantom{\kern-.5\wickgap\copy\wickbox}}
 \kern-\wd\wickbox\kern.5\wickgap
 \rlap{\copy\wickbox}
 \c#3{\phantom{\kern.5\wickgap\box\wickbox}}
 \kern-.5\wickgap}

\title{Error Scaling of Sea Quark Isospin-Breaking Effects}

\author[a]{Anian Altherr}
\author[b]{Isabel Campos}
\author*[c,d]{Alessandro Cotellucci}
\author[a]{Roman Gruber}
\author[a]{Tim Harris}
\author[a]{Marina Krstić Marinković}
\author[a]{Letizia Parato}
\author[c,e]{Agostino Patella}
\author[c,b]{Sara Rosso}
\author[a]{Paola Tavella}

\affiliation[a]{Institut für Theoretische Physik, ETH Zürich, Wolfgang-Pauli-Str. 27, 8093 Zürich, Switzerland}
\affiliation[b]{Instituto de Física de Cantabria and IFCA-CSIC, Avda. de Los Castros s/n, 39005 Santander, Spain}
\affiliation[c]{Humboldt Universität zu Berlin, Institut für Physik and IRIS Adlershof, Zum Großen Windkanal 6, 12489
Berlin, Germany}
\affiliation[d]{Jülich Supercomputing Centre, Forschungszentrum Jülich, D-52428 Jülich, Germany}
\affiliation[e]{DESY, Platanenallee 6, D-15738 Zeuthen, Germany}

\emailAdd{a.cotellucci@fz-juelich.de}

\abstract{Sea-quark isospin-breaking effects (IBE) are difficult to compute since they require the evaluation of all-to-all propagators. However, the quest for high-precision calculations motivates a detailed study of these contributions. There are strong arguments that the stochastic error associated with these quantities should diverge in the continuum and infinite-volume limit, resulting in a possible bottleneck for the method. In this work, we present the study of the error scaling for these quantities using $N_f=3$ $O(a)$-improved Wilson fermions QCD with C-periodic boundary conditions in space, a pion mass $M_{\pi}=400$ MeV, a range of lattice spacings $a=0.05, 0.075, 0.1$ fm, and volumes $L=1.6, 2.4, 3.2$ fm. The analysis of the error as a function of the number of stochastic sources shows that we reach the gauge error for the dominant contributions. The errors do not show the leading order divergence $1/a$ for strong-IBE and $1/a^2$ for electromagnetic IBE, in the considered range of lattice spacings. On the other hand, our data are consistent with the predicted leading divergence $\sqrt{V}$.}

\FullConference{The 41st International Symposium on Lattice Field Theory (LATTICE2024)\\
 28 July - 3 August 2024\\
Liverpool, UK\\}

\begin{document}
\maketitle

\section{Motivations}
Isospin symmetry is a good approximation for Hadronic Phenomena as long as the target precision is above $1\%$ otherwise electromagnetic and strong isospin-breaking effects (IBE) need to be taken into account. In fact, the up-down mass difference and QED coupling constant are both of order $O\left(0.01\right)$:
\begin{equation}
\underbrace{\left(\frac{m_d-m_u}{\Lambda_{\text{QCD}}}\right) \sim 0.01}_{\text{Strong-IBE}}\ \hspace{2cm} \underbrace{\frac{e^2}{4\pi} \sim 0.01}_{\text{Electromagnetic-IBE}}.
\end{equation}
The evaluation of IBE effects in lattice simulations can be done by reweighing of isosymmetric QCD simulations and perturbative expansion in $(m_u-m_d)$ and $e^2$ namely the RM123 method \citep{deDivitiis:2013xla}.

The QCD+QED action is rewritten as $S_{\text{QCD+QED}}=S_{\text{Iso}+\gamma}+S_{\text{IB}}\left(\delta m,e^2\right)$ and the path integral average of an observable $\mathcal{O}$ is expanded as:
\begin{equation}
\begin{aligned}
\langle\mathcal{O}\rangle =&\langle\wick{\self{2}{\text{$\mathcal{O}$}}}\rangle_{\text{Iso}}-\underbrace{\langle\wick{\c1{S}_{\text{IB}} \c1{\mathcal{O}}} \rangle_{\text{Iso}\gamma,c}+\frac{1}{2}\langle \wick{\c1{S}_{\text{IB}} \c2{S}_{\text{IB}} \doublewick\c1{\mathcal{O}}\c2}\rangle_{\text{Iso}\gamma,c}}_{\text{valence-valence}}+\underbrace{\langle \wick{\self{2}{S}_{\text{IB}}} \wick{\c1{S}_{\text{IB}}\c1{\mathcal{O}}}\rangle_{\text{Iso}\gamma,c}}_{\text{sea-valence}}\\
&-\underbrace{\langle\wick{\self{2}{S}_{\text{IB}}} \wick{\self{2}{\text{$\mathcal{O}$}}} \rangle_{\text{Iso}\gamma,c}+\frac{1}{2}\langle \wick{\self{2}{S}_{\text{IB}}} \wick{\self{2}{S}_{\text{IB}}} \wick{\self{2}{\text{$\mathcal{O}$}}}\rangle_{\text{Iso}\gamma,c}+\frac{1}{2}\langle \wick{\c1{S}_{\text{IB}} \c1{S}_{\text{IB}}} \wick{\self{2}{\text{$\mathcal{O}$}}}\rangle_{\text{Iso}\gamma,c}}_{\text{sea-sea}}+O \left(\delta m_{ud}^2,e^4,e^2\delta m_{ud}\right)
\end{aligned}
\end{equation}
the different Wick contractions are referred to by different names: valence-valence, sea-valence and sea-sea. The sea-sea contractions are usually neglected since they are expected to be small and noisy. However they need to be estimated in order to have a full error budget. Other works in this conference addressed this problem \citep{Hill:2024}. This work focuses on the sea-sea diagrams which are:
\begin{equation}
\begin{aligned}
\langle\mathcal{O}\rangle_{\text{sea}}=&\sum_f\delta m_f\langle \underset{\text{mass}}{\vcenter{\hbox{\begin{tikzpicture}[>={Triangle[bend,width=2pt,length=3pt]}]
        \coordinate (z);
        \node at (z)[circle,fill,inner sep=0.6pt]{};
  \draw[decoration={markings,mark=at position 0.25 with {\coordinate (B);}, mark=at position 0.75 with {\coordinate (C);}},postaction={decorate}] (z) arc[start angle=0,end angle=360,radius=0.25];
		\node at (B)[label={[label distance=-0.2cm]90:\tiny{f}}]{};
       	\node at (C)[label={[label distance=-0.05cm]270:\tiny{}}]{};
        \draw[->] (z) arc[start angle=0,end angle=200,radius=0.25];
\end{tikzpicture}}}}\ \mathcal{O}\rangle_{\text{Iso},c}+e^2\sum_fq_f^2 \langle\underset{\text{tadpole}}{\vcenter{\hbox{\begin{tikzpicture}[>={Triangle[bend,width=2pt,length=3pt]}]
        \coordinate (z);
        \coordinate[below left = 0.0094 and 0.033 of z] (w);
        \node at (z)[circle,fill,inner sep=0.6pt]{};
        \draw[decoration=Snake,segment length=2.9342574999999pt,segment amplitude=0.9pt, decorate] (w) arc[start angle=-180,end angle=180,radius=0.14999878];
        \draw[decoration={markings,mark=at position 0.25 with {\coordinate (B);}, mark=at position 0.75 with {\coordinate (C);}},postaction={decorate}] (z) arc[start angle=0,end angle=360,radius=0.25];
\node at (B)[label={[label distance=-0.2cm]90:\tiny{f}}]{};
       	\node at (C)[label={[label distance=-0.05cm]270:\tiny{}}]{};
        \draw[->] (z) arc[start angle=0,end angle=200,radius=0.25];
\end{tikzpicture}}}}\ \mathcal{O}\rangle_{\text{Iso},c}\\
&+e^2\left[\underset{\text{lightbulb}}{\sum_fq_f^2\langle\vcenter{\hbox{\begin{tikzpicture}[>={Triangle[bend,width=2pt,length=3pt]}]
        \coordinate (z);
        \coordinate[below=0.5 of z] (w);
        \node at (w)[circle,fill,inner sep=0.6pt, label={[label distance=0.05cm]270:\tiny{}}]{};
        \node at (z)[circle,fill,inner sep=0.6pt, label={[label distance=-0.1cm]90:\tiny{f}}]{};
        \draw (z) arc[start angle=90,end angle=450,radius=0.25];
        \draw[->] (z) arc[start angle=90,end angle=200,radius=0.25];
        \draw[->] (w) arc[start angle=270,end angle=380,radius=0.25];
        \draw[decoration=snake,segment length=2.9pt,segment amplitude=0.9pt,decorate] (z) -- (w);
\end{tikzpicture}}}\ \mathcal{O}\rangle_{\text{Iso},c}}+\sum_{fg}q_fq_g\langle\underset{\text{lanterns}}{\vcenter{\hbox{\begin{tikzpicture}[>={Triangle[bend,width=2pt,length=3pt]}]
        \coordinate (z);
        \coordinate[right=0.75 of z] (w);
        \node at (w)[circle,fill,inner sep=0.6pt]{};
        \node at (z)[circle,fill,inner sep=0.6pt]{};
        \draw[decoration={markings,mark=at position 0.25 with {\coordinate (B);}, mark=at position 0.75 with {\coordinate (C);}},postaction={decorate}]  (z) arc[start angle=0,end angle=360,radius=0.25];
        \draw[decoration={markings,mark=at position 0.25 with {\coordinate (D);}, mark=at position 0.75 with {\coordinate (E);}},postaction={decorate}]  (w) arc[start angle=-180,end angle=180,radius=0.25];
		\node at (B)[label={[label distance=-0.2cm]90:\tiny{f}}]{};
       	\node at (C)[label={[label distance=-0.05cm]270:\tiny{}}]{};
		\node at (E)[label={[label distance=-0.2cm]90:\tiny{g}}]{};
       	\node at (D)[label={[label distance=-0.05cm]270:\tiny{}}]{};
        \draw[->] (z) arc[start angle=0,end angle=200,radius=0.25];
        \draw[->] (w) arc[start angle=180,end angle=380,radius=0.25];
        \draw[decoration=snake,segment length=2.9pt,segment amplitude=0.9pt,decorate] (z) -- (w);
\end{tikzpicture}}}}\ \mathcal{O}\rangle_{\text{Iso},c}\right].
\end{aligned}
\end{equation}
where the $\delta m_f$s need to be tuned according to a renormalization scheme.

The following study has been performed using $N_f=3$  $O(a)$-improved Wilson fermions with C$^{\star}$ boundary conditions in space
and periodic boundary conditions in time, the bare action parameters have been taken from the CLS ensembles \cite{RQCD:2022xux}. The SU(3) flavour symmetry provides some simplifications to the diagrammatic expansion since the total sea-valence and the sea-sea lanterns diagrams are zero. We compute the sea-sea diagrams using a two-level frequency splitting \citep{Giusti:2019kff} and hopping expansion \cite{Thron_1998} at the charm mass, tuned by \citep{Ce:2022kxy}.

\section{Scaling of the gauge error}

The statistical error of the insertion of each diagram $\mathcal{D}$ in the expectation value of some observable $\mathcal{O}$ can be studied assuming independence of the gauge configurations and Gaussian random sources, as already done in \cite{Harris:2023zsl, Harris:2024dky}. The variance of such insertions can be written as the sum of different contributions: the gauge variance due to the use of statistically sampled gauge configurations, the source variance due to the use of random sources to compute the trace, the photon variance due to the use of stochastic photons to compute the photon propagator. If enough stochastic sources and photons are used, the error of the estimator is dominated by the gauge noise.

Using standard techniques, the gauge variance can be expressed as integrals of certain $n$-point functions in a partially-quenched extension of QCD. The scaling towards the continuum limit $a\rightarrow 0$ can be studied by means of the (off-shell) Symanzik effective theory and the operator product expansion. The scaling towards the infinite volume limit $V \to \infty$ can be studied using a cluster decomposition the $n$-point functions, translational invariance and the (far from obvious) assumption that connected expectation values decay exponentially fast with the distance in the partially-quenched theory. Details will be given in a future publication. One finds that, at the leading order in either of the $a\rightarrow 0$ and $V \to \infty$ limits and assuming local observables in spacetime, the gauge noise of any insertion factorizes as in:
\begin{equation}
\sigma\left[\langle\mathcal{D}\mathcal{O}\rangle_{\text{Iso},c}\right]
\sim
\sigma_{\mathcal{D}} \sigma_{\mathcal{O}} ,
\end{equation}
where $\sigma_{\mathcal{D}}$ and $\sigma_{\mathcal{O}}$ denote the statistical errors of the expectation values $\langle\mathcal{D}\rangle_{\text{Iso}}$ and $\langle\mathcal{O}\rangle_{\text{Iso}}$, respectively. Moreover one finds that the insertion of the strong-IB diagram $\mathcal{D}_{SIB}$ scales as
\begin{equation}
\sigma\left[\langle\mathcal{D}_{SIB}\mathcal{O}\rangle_{\text{Iso},c}\right]
\sim
\sigma_{\mathcal{O}}
a^{-1}\sqrt{V}
,
\end{equation}
while the gauge noise of the insertion of any of the electromagnetic-IB diagrams $\mathcal{D}_{EIB}$ scales as
\begin{equation}
\sigma\left[\langle\mathcal{D}_{EIB}\mathcal{O}\rangle_{\text{Iso},c}\right]
\sim
\sigma_{\mathcal{O}}a^{-2}\sqrt{V}
.
\end{equation}
Notice that the electromagnetic diagrams present a more severe divergence of the gauge error with the lattice spacing compared to the strong-IB diagram.

\section{Volume scaling}
The volume scaling of the gauge error of the diagrams has been investigated using the ensembles presented in Table \ref{tab:phys_val_vol}, which cover the range $1.6\ \text{fm} <L<3.2\ \text{fm}$.

\begin{table}[h!]
    \small
    \centering
    \setlength{\columnsep}{-1.5cm}
    \addtolength{\leftskip}{-2cm}
    \addtolength{\rightskip}{-2cm}
    \begin{tabular}{cccccc}
        \hline
        ensemble & geometry & $a$ [fm] &
        $M_{\pi}$ [MeV] & $M_{\pi}L$ &  n.cnfg \\
        \hline
        \texttt{A420a00b334} & $32\times 16^3$ & $0.0990(7)$ & $413(8)$ & $3.31$ & $50$  \\
        \texttt{B420a00b334} & $48\times 24^3$  & $0.0991(3)$ & $415(3)$ & $4.98$ & $50$ \\
        \texttt{C420a00b334} & $64\times 32^3$ & $0.0990(1)$ & $415(1)$ & $6.63$ & $50$ \\
        \hline
    \end{tabular}
    \caption{Ensembles for the volume scaling. The $50$ gauge configurations are independent.}
   \label{tab:phys_val_vol}
\end{table}

In Figure \ref{tab:phys_val_vol}, the error on the expectation value of individual sea-sea diagram is plotted as a function of the number of stochastic sources up to 400, for each ensemble. The mass and the tadpole diagrams reach the gauge noise already for $O\left(10\right)$ random sources. The lightbulb diagrams involving the QED SW term reach the gauge for $O\left(50\right)$ random sources, while the lightbulb diagram constructed only with the electromagnetic current does not reach the gauge noise in the considered range, however its contribution to the error becomes negligible compared to the others. Similarly, the lanterns diagrams do not reach the gauge noise, however their contribution to the error becomes negligible for $O\left(200\right)$ random sources.

\begin{figure}[h!]
\centering
\includegraphics[width=10cm]{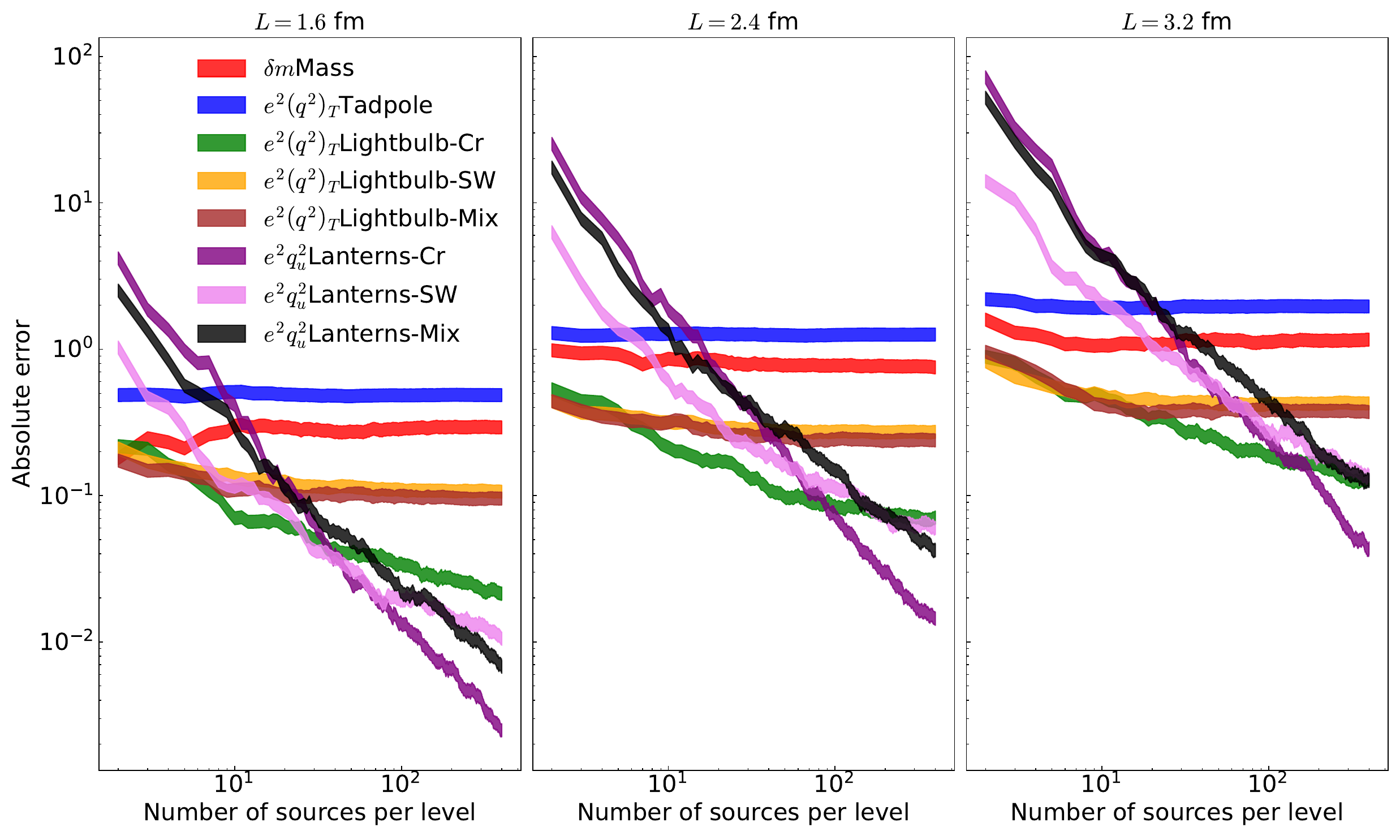}
\caption{Scaling of the absolute error of the sea-sea diagrams for different random sources for the three different volumes analysed. The mass diagram is multiplied times the total mass shift $\delta m$ tuned to absorb the radiative divergences but the error on $\delta m$ is not propagated.}
\label{fig:Diagrams_scale_vol}
\end{figure}

\begin{figure}[h!]
  \centering
  \begin{tabular}{cc}
\includegraphics[width=1.65in]{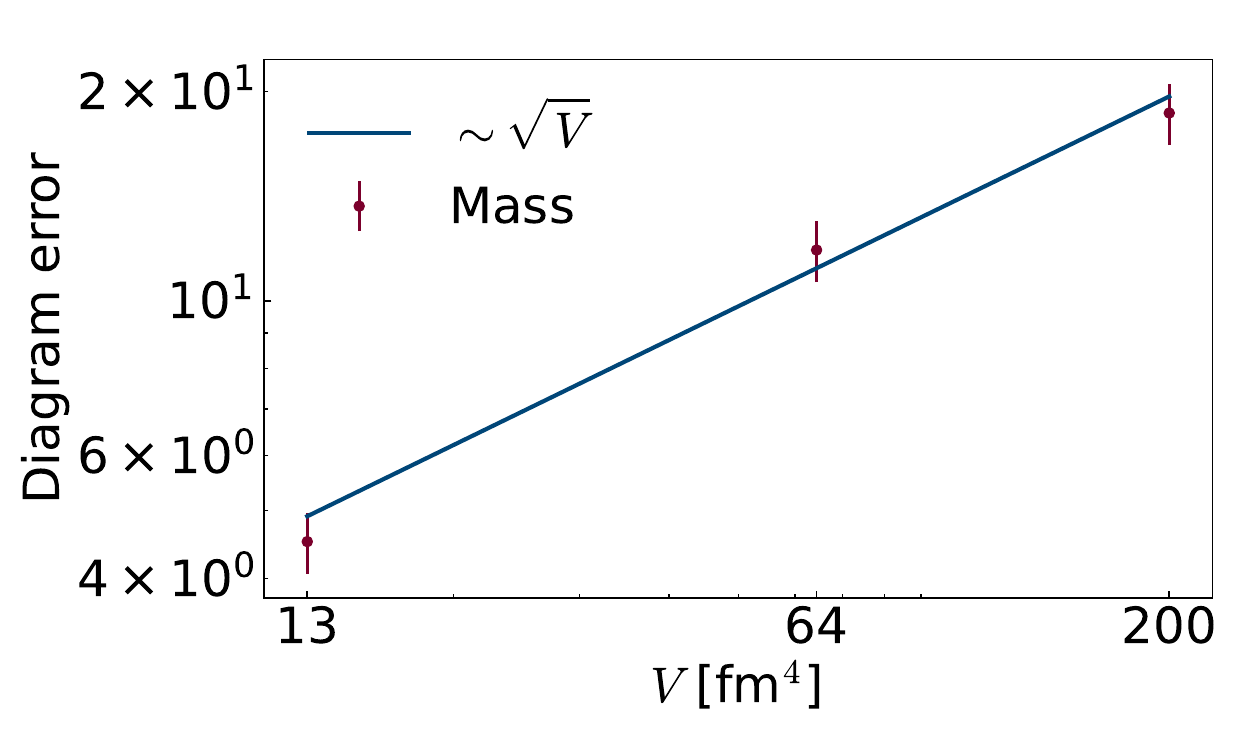} &
\includegraphics[width=1.65in]{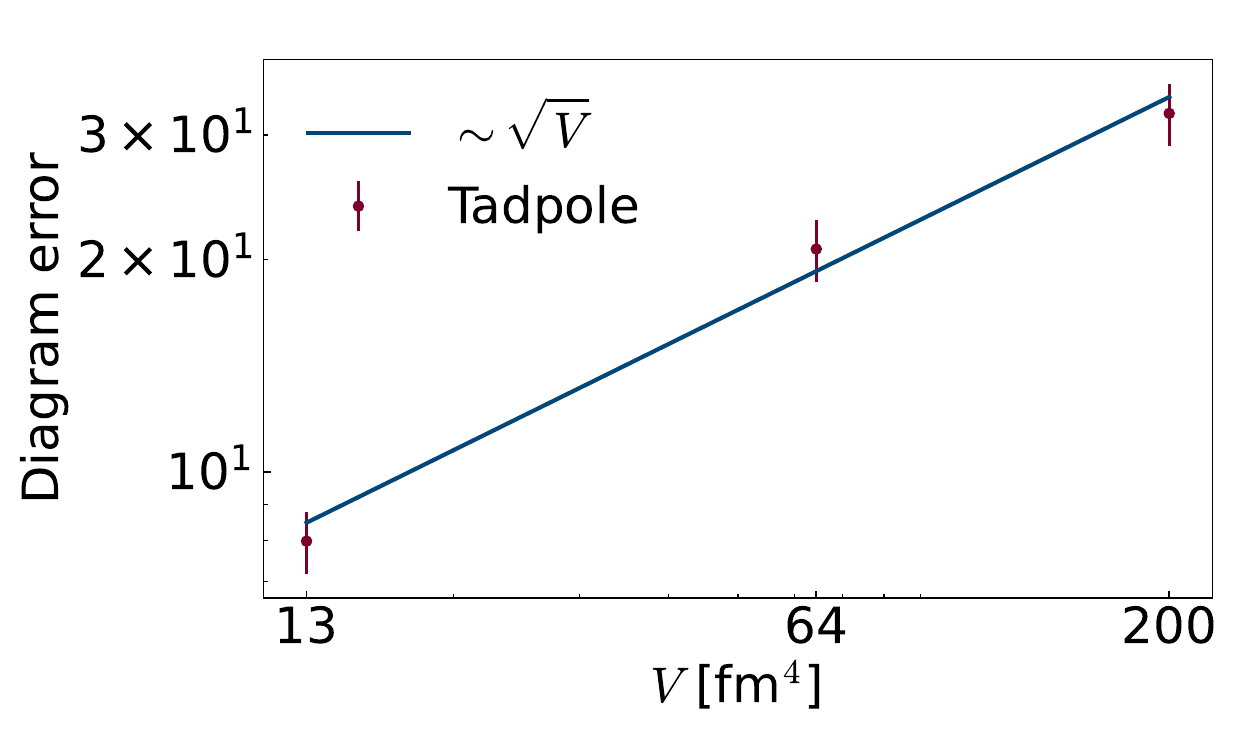}\\
\end{tabular}\\
     \addtolength{\leftskip}{-3cm}
    \addtolength{\rightskip}{-3cm}
    \setlength{\tabcolsep}{-0.02cm}
 \begin{tabular}{ccc}
\includegraphics[width=1.65in]{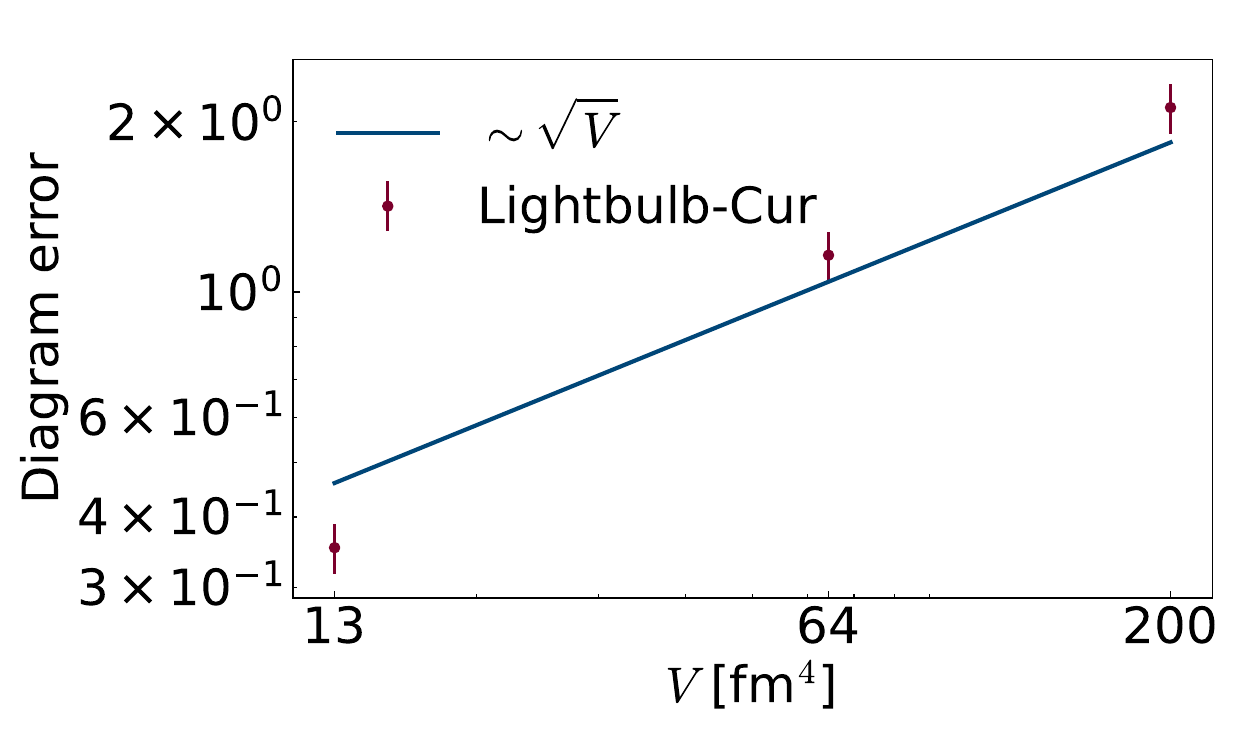} &
\includegraphics[width=1.65in]{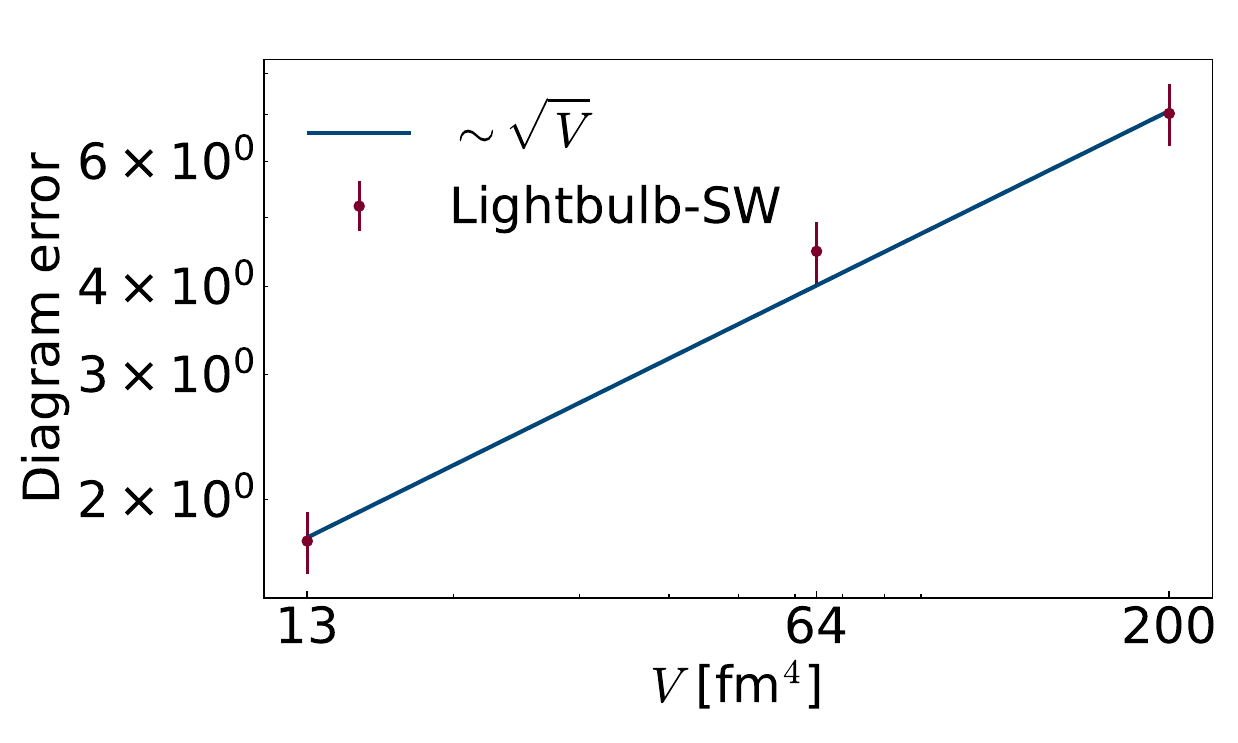} &
\includegraphics[width=1.65in]{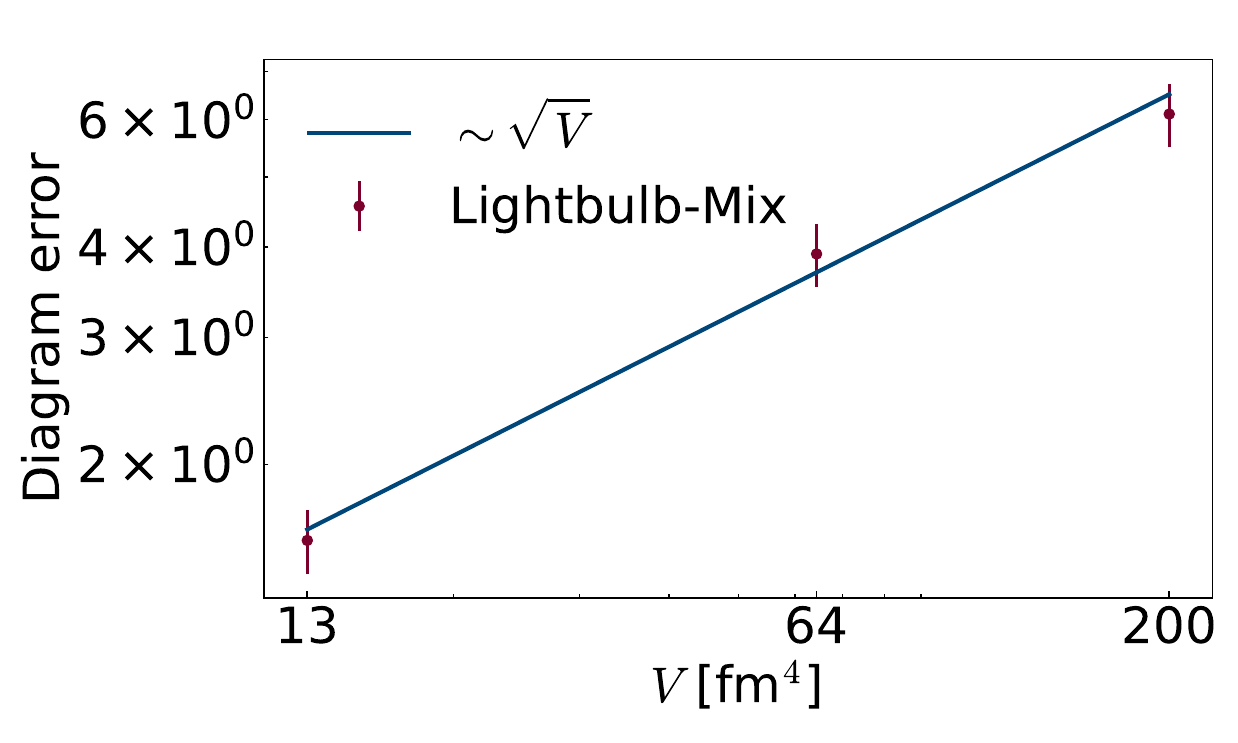}
\end{tabular}
\caption{Volume scaling of the error of the single diagrams in logarithmic scale.}
\label{fig:err_scale_vol}
\end{figure}

From Figure \ref{fig:Diagrams_scale_vol} we can observe the increase of the absolute error going from small to big volumes and Figure \ref{fig:err_scale_vol} shows the scaling of the errors for different volumes for the physically relevant diagrams. All the data are compatible with the expected $\sqrt{V}$ scaling of the error.

\section{Continuum scaling}
Table \ref{tab:phys_val_a} shows the ensembles used to investigate the scaling of the error of the sea-sea diagrams towards the continuum limit. The ensembles include lattice spacings in the range $0.05\ \text{fm} <a<0.10\ \text{fm}$.

\begin{table}[h!]
    \small
    \centering
    \setlength{\columnsep}{-1.5cm}
    \addtolength{\leftskip}{-2cm}
    \addtolength{\rightskip}{-2cm}
    \begin{tabular}{cccccc}
        \hline
        ensemble & geometry & $a$ [fm] &
        $M_{\pi}$ [MeV] &  $L$ [fm] & n.cnfg \\
        \hline
        \texttt{C420a00b370}  & $64\times 32^3$ & $0.0499(2)$ & $416(3)$ & $1.596(6)$ & $50$  \\
        \texttt{B420a00b346} & $48\times 24^3$  & $0.0769(3)$ & $414(2)$ & $1.857(6)$ & $50$  \\
        \texttt{A420a00b334}  & $32\times 16^3$ &  $0.0990(7)$ & $413(8)$ & $1.58(1)$  & $50$ \\
        \hline
    \end{tabular}
     \caption{Ensembles used for the continuum scaling. The $50$ gauge configurations are independent.}
   \label{tab:phys_val_a}
\end{table}

\begin{figure}[h!]
\centering
\includegraphics[width=10cm]{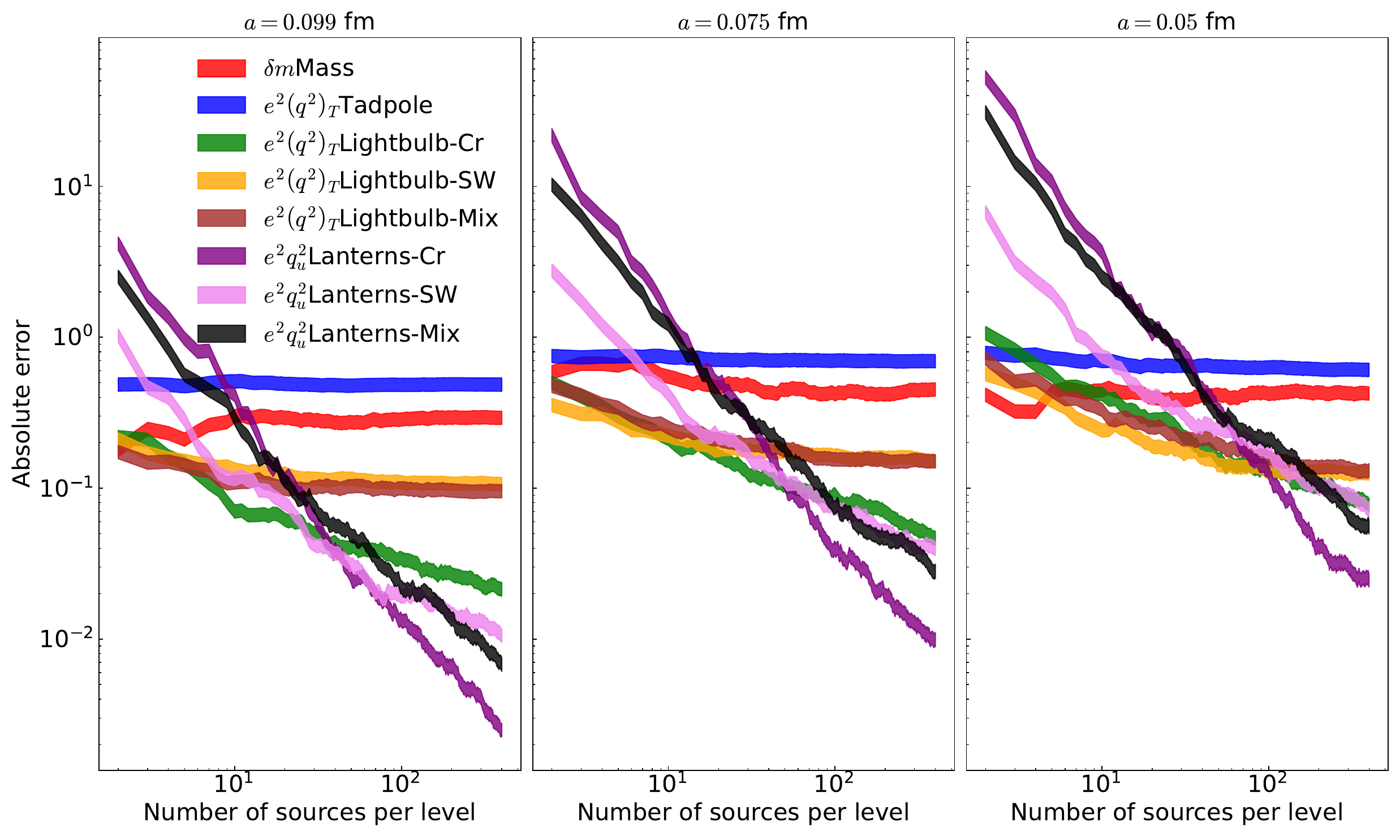}
\caption{Scaling of the absolute error of the sea-sea diagrams for different random sources for the three different lattice spacings analysed. The mass diagram is multiplied times the total mass shift $\delta m$ tuned to absorb the radiative divergences but the error on $\delta m$ is not propagated.}
\label{fig:Diagrams_scale_a}
\end{figure}

In Figure \ref{fig:Diagrams_scale_a}, the error on the expectation value of individual sea-sea diagram is plotted as a function of the number of stochastic sources up to 400, for each ensemble.

As for the other ensembles, the mass and the tadpole diagrams are the dominant error sources and they reach the gauge error around $O\left(10\right)$ sources per level. The lightbulb diagrams, instead, reach the gauge noise only for the ones involving the U(1) SW operator but we can see that the starting point of the plateau shifts to a higher number of sources going to a finer lattice spacing. The lightbulb involving only the current does not reach the gauge error but it becomes irrelevant compared to the other errors. None of the lanterns diagrams reach the gauge error and we can see that the source error increases going to a finer lattice spacing.

\begin{figure}[h!]
  \centering
     \addtolength{\leftskip}{-3cm}
    \addtolength{\rightskip}{-3cm}
    \setlength{\tabcolsep}{-0.02cm}
  \begin{tabular}{cc}
\includegraphics[width=2.0in]{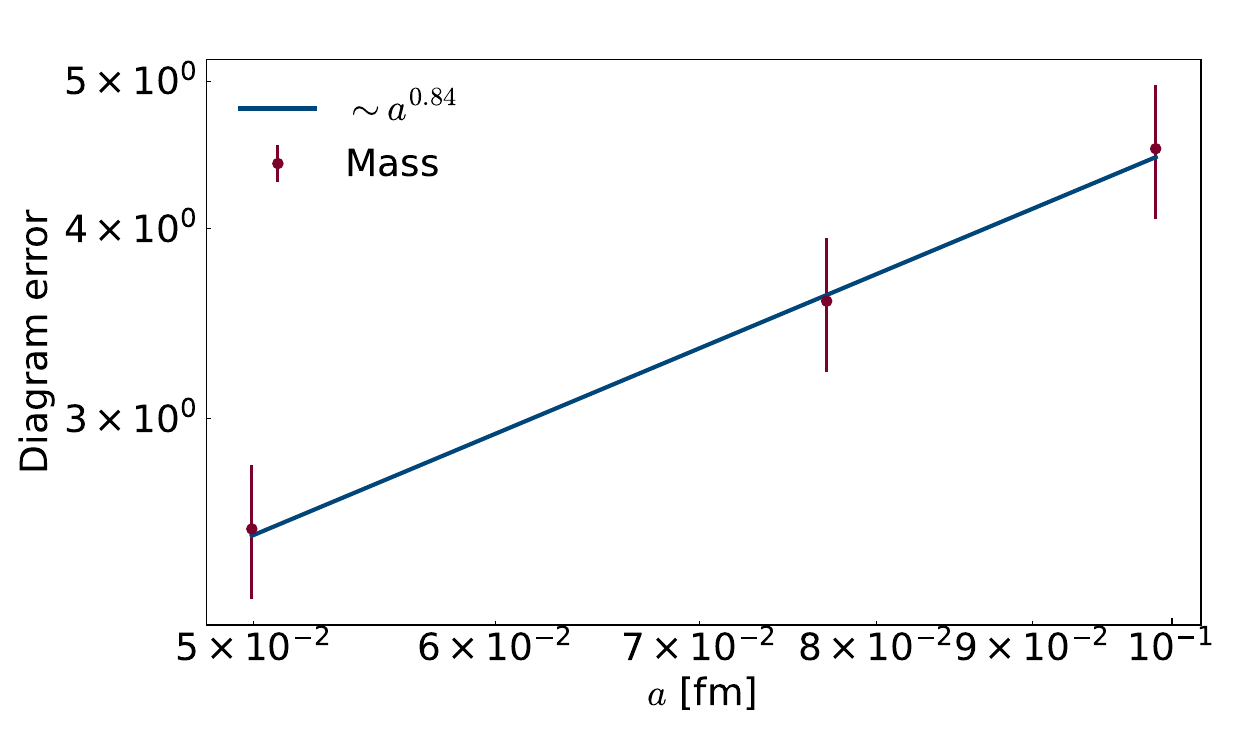} & \includegraphics[width=2.0in]{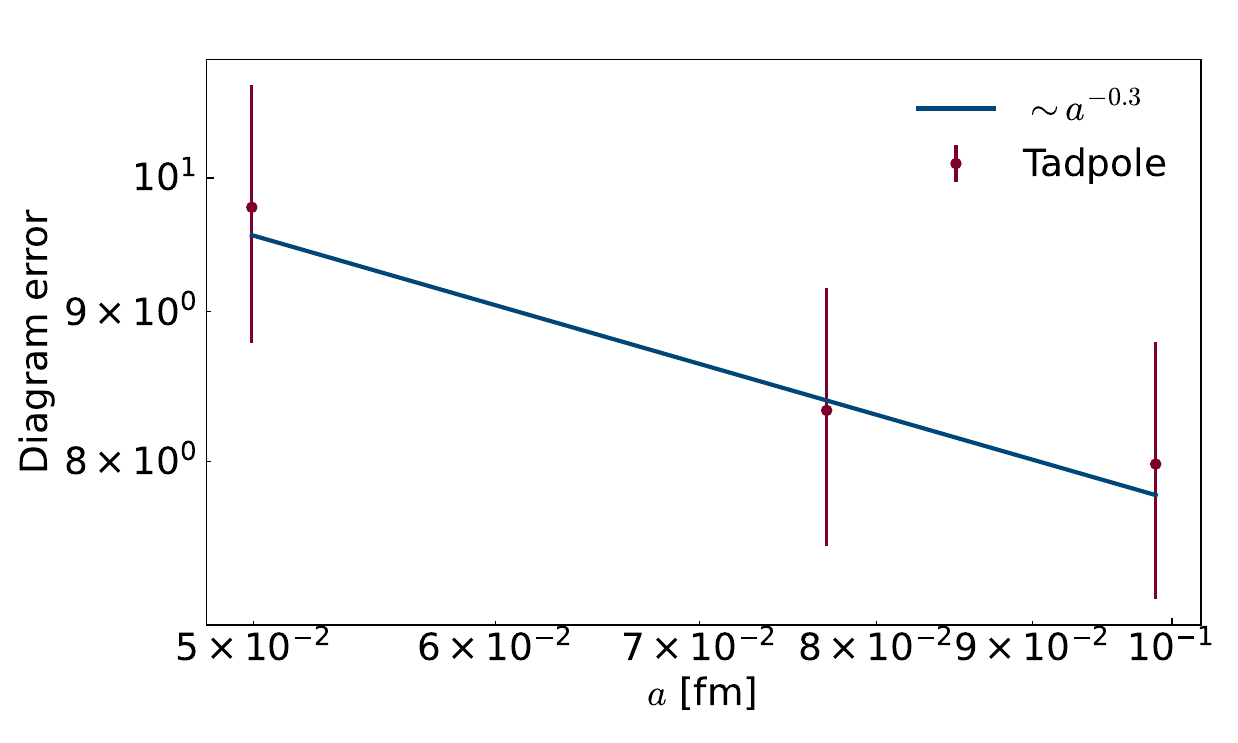}\\
\includegraphics[width=2.0in]{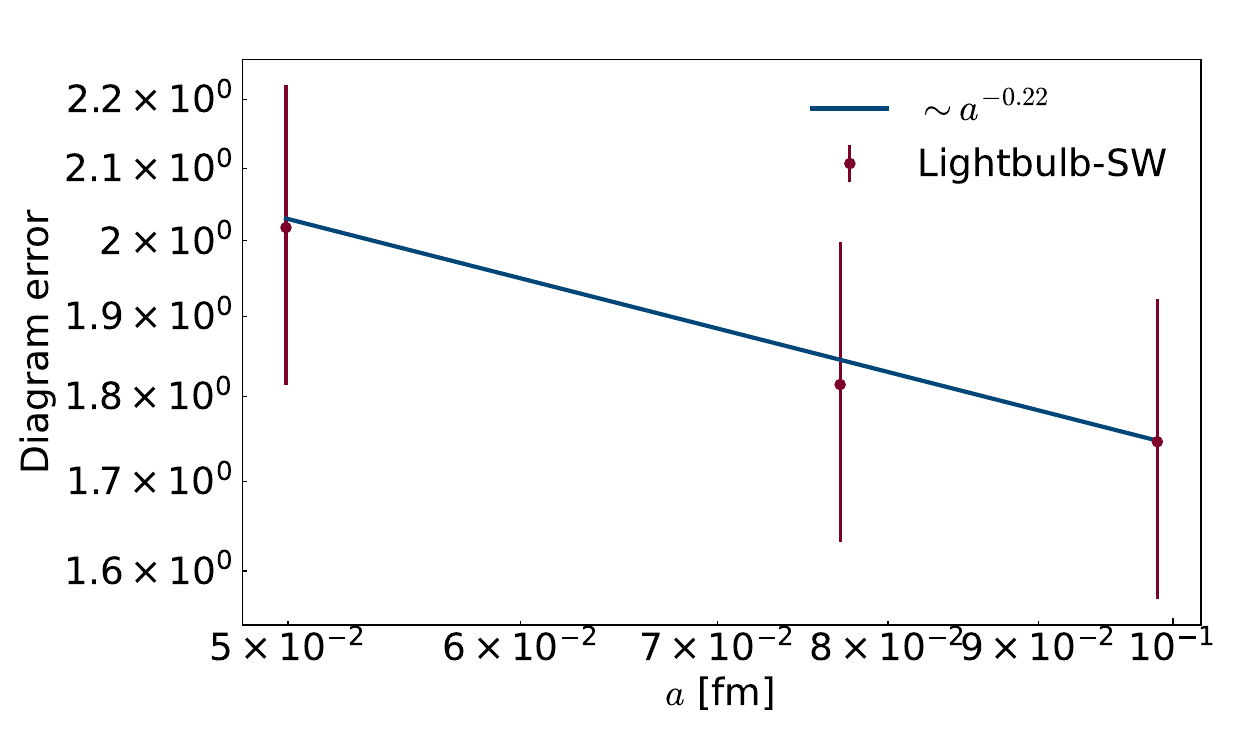} &
\includegraphics[width=2.0in]{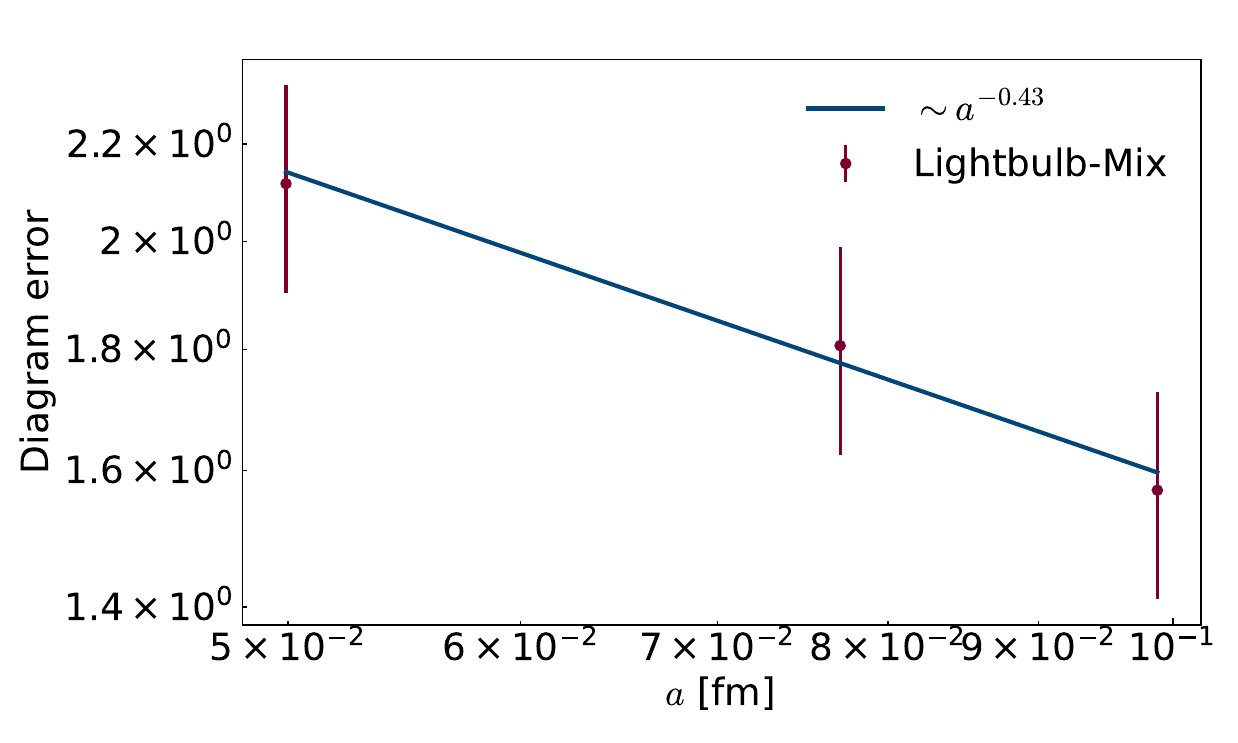}
\end{tabular}
\caption{Continuum scaling of the error of the single diagrams in logarithmic scale. To account for the small differences in volume between the different ensembles the error, of each ensemble $i$, has been rescaled by a factor $\sqrt{V^{\text{\texttt{A420a00b334}}}}/\sqrt{V_i}$, a power-law fit of the data is also included to see the power of the scaling.}
\label{fig:err_scale_a}
\end{figure}

Figure \ref{fig:err_scale_a} shows the gauge error scaling for the diagrams for different lattice spacings. The asymptotic divergence is not observed in the considered ensembles. By fitting the data to a power law, we see that the error increases with a lower power of the lattice spacing in this range (in the case of the mass diagram, the error even decreases).

\section{Effect on $t_{0}/a^2$}

\begin{figure}[h!]
\centering
\includegraphics[width=10cm]{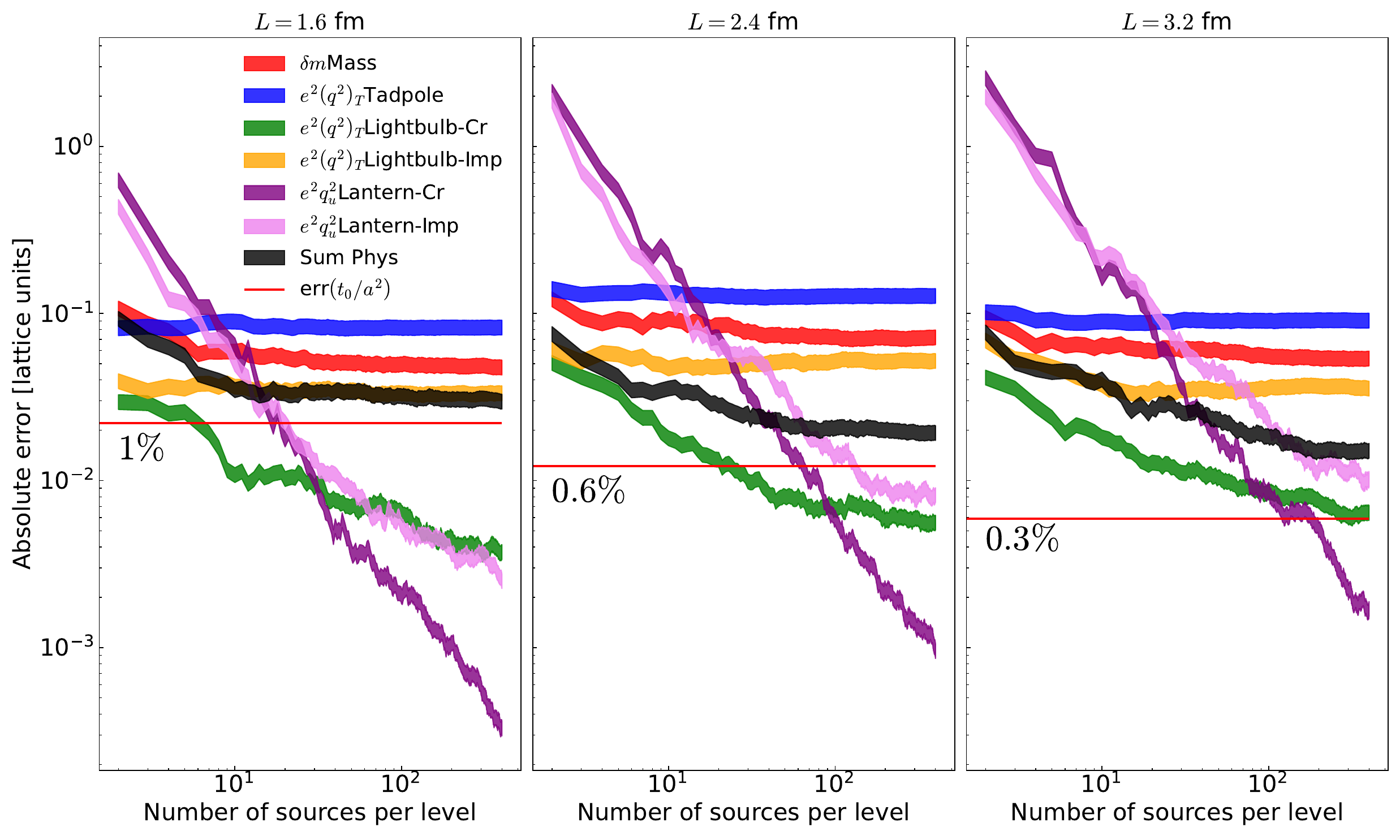}
\caption{Scaling of the absolute error of the sea-sea diagrams contribution to $t_{0}/a^2$ for different random sources for the three different volumes analysed. The mass diagram is multiplied times the total mass shift $\delta m$ tuned to absorb the radiative divergences and the error on $\delta m$ is propagated. The horizontal line marks the absolute error on the isosymmetric part of $t_{0}/a^2$ and the relative precision is written as a label.}
\label{fig:t0_scale_vol}
\end{figure}

\begin{figure}[h!]
\centering
\includegraphics[width=10cm]{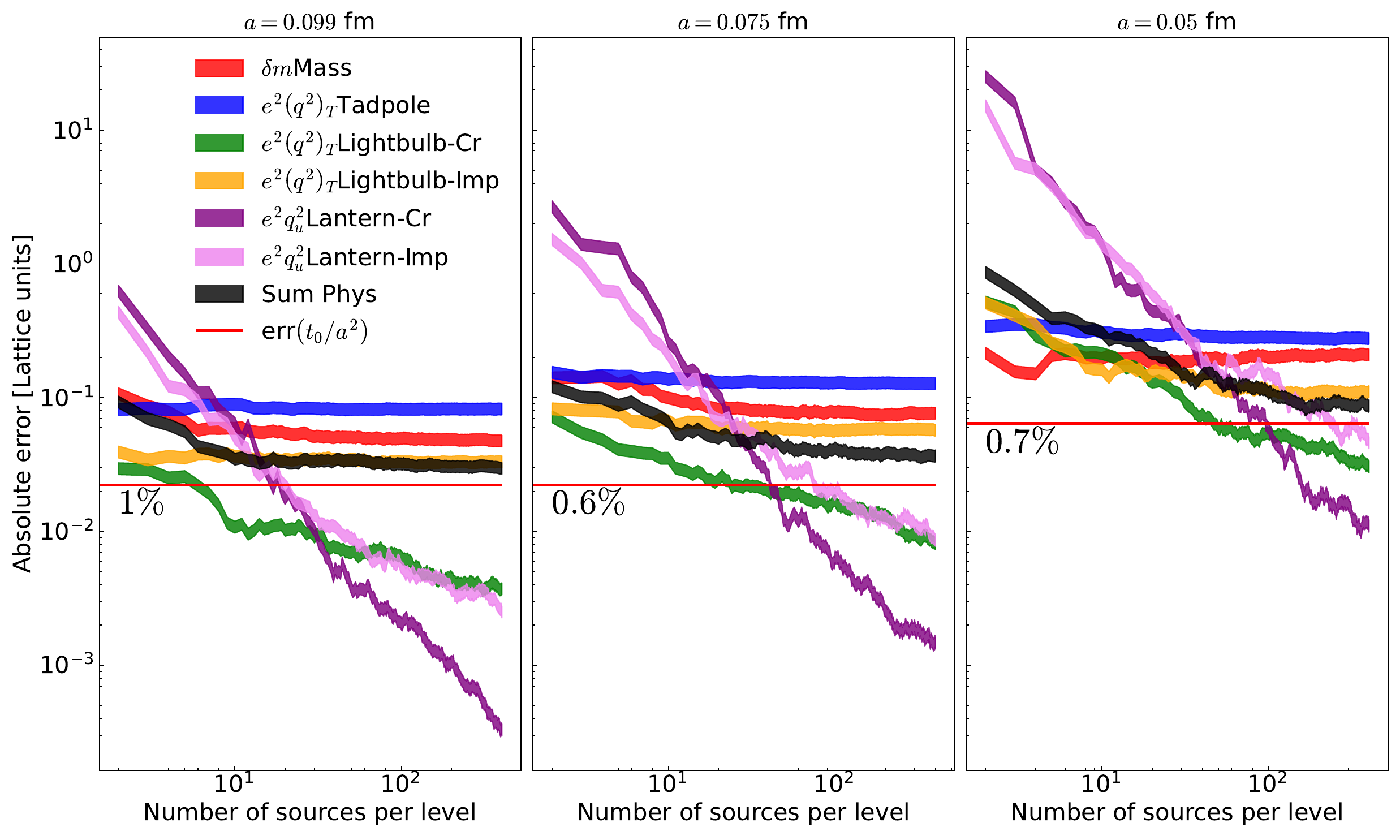}
\caption{Scaling of the absolute error of the sea-sea diagrams contribution to $t_{0}/a^2$ for different random sources for the three different lattice spacings analysed. The mass diagram is multiplied times the total mass shift $\delta m$ tuned to absorb the radiative divergences and the error on $\delta m$ is propagated. The horizontal line marks the absolute error on the isosymmetric part of $t_{0}/a^2$ and the relative precision is written as a label.}
\label{fig:t0_scale_a}
\end{figure}

We extend the study of the IBE contributions to the Wilson-flow-based scale $t_{0}$. Figures \ref{fig:t0_scale_vol} and \ref{fig:t0_scale_a} show the error scaling with the number of sources of the diagrams insertions to $t_{0}/a^2$ for the different ensembles. The plots include the error of the physical sum of the insertions and flat lines representing the error on the isoQCD determination of $t_{0}/a^2$. We see from the plots that the scaling is similar to the scaling of the error of the diagrams themselves corroborating the factorization formula. For all the analysed ensembles, the total statistical error on the sea-sea isospin breaking corrections to $t_{0}/a^2$ is bigger than the stochastic error on the isosymmetric QCD part of $t_{0}/a^2$.

\begin{figure}[h!]
\centering
\includegraphics[width=8cm]{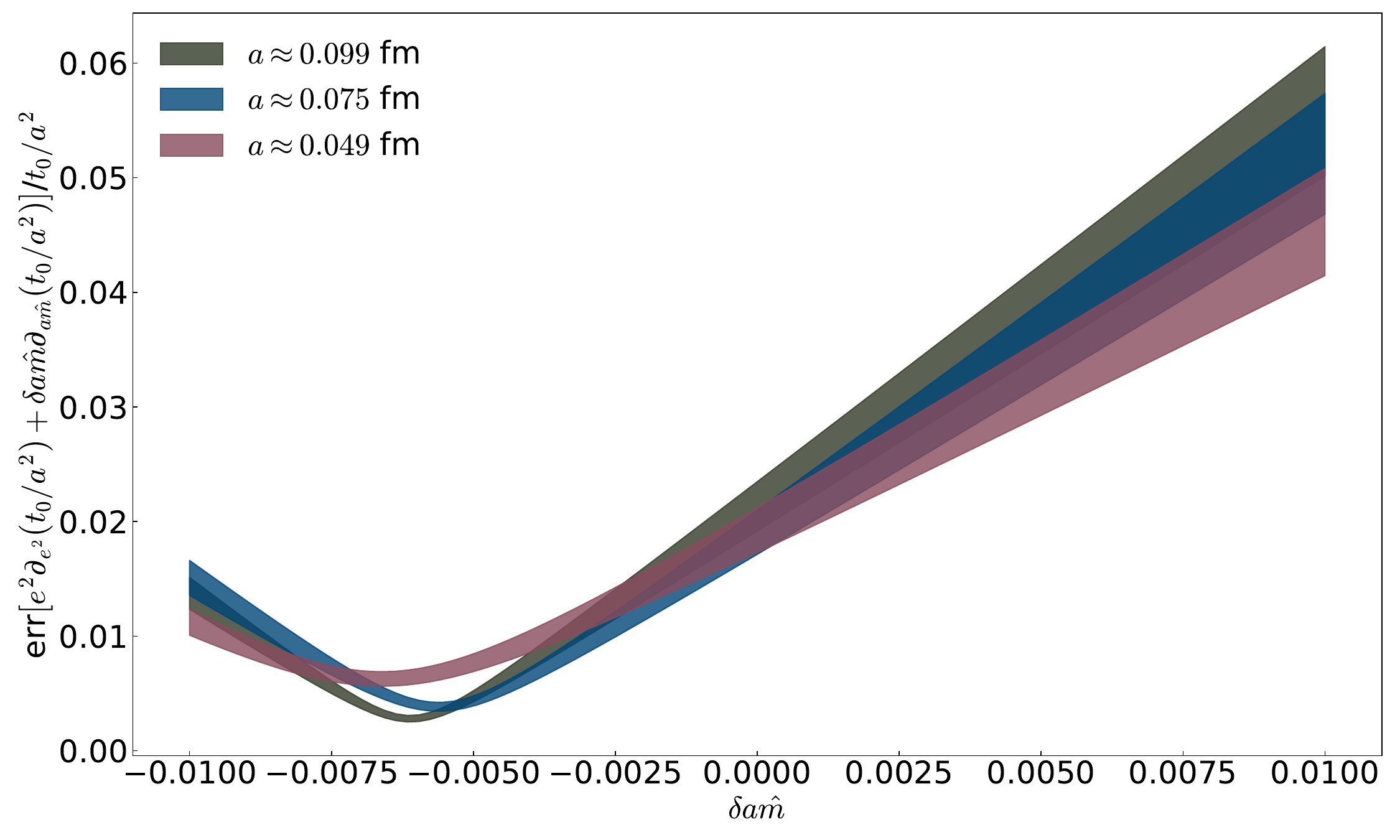}
\caption{Error of the total IBE correction to $t_{0}/a^2$ divided by the isosymmetric QCD value of $t_{a}/a^2$, as a function of the total bare quark mass shift $\delta m$ (in lattice units) for the three different lattice spacings considered.}
\label{fig:Correlation}
\end{figure}

Figure \ref{fig:Correlation} shows the behaviour of the error of the total IBE correction to $t_{0}/a^2$ divided by the isosymmetric QCD value of $t_{0}/a^2$, as a function of the total bare quark mass shift $\delta m$ (in lattice units) for the different lattice spacings analysed. The error shows a minimum for a negative value of the total mass shift which implies that including mass renormalization has the effect to decrease the error on the sea-sea contributions. The cancellation between the mass counter term and the leading electromagnetic diagram, the tadpole can be made exact by choosing a point-split discretization of the mass operator.

\begin{figure}[h!]
\centering
\includegraphics[width=7.5cm]{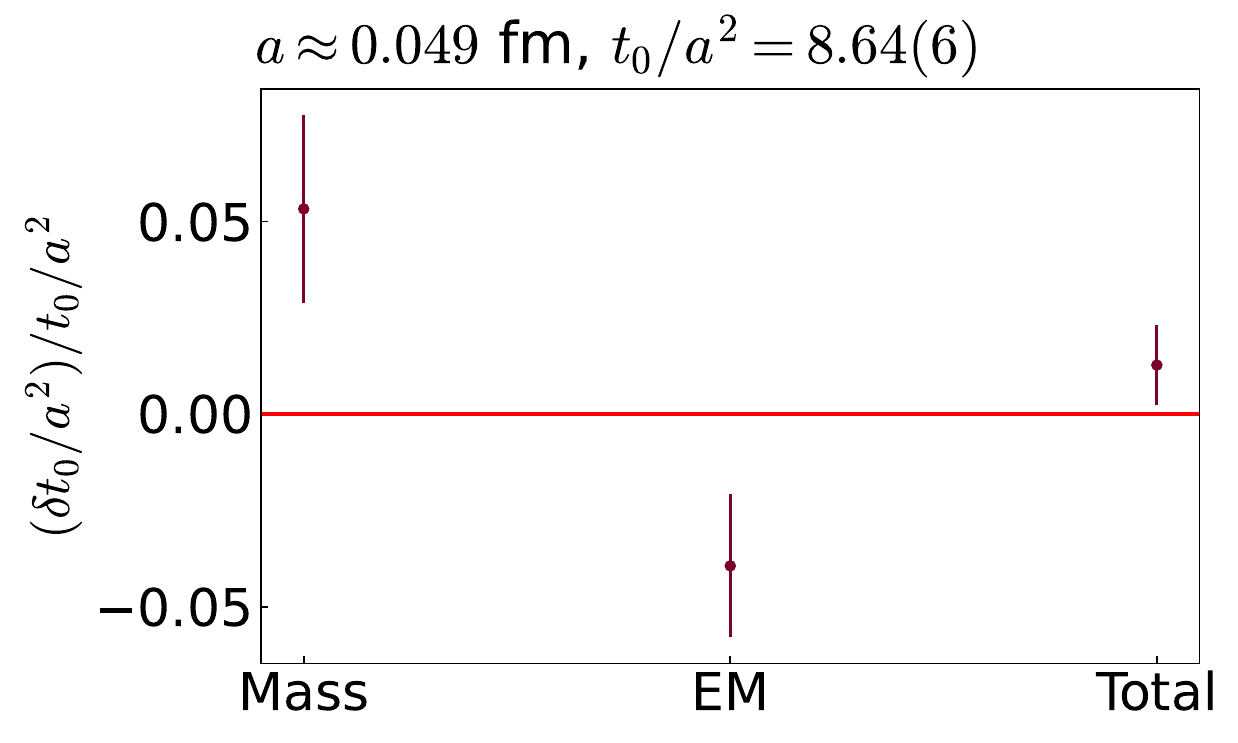}
\caption{The isospin breaking effects to $t_{0}/a^2$ divided by the isosymmetric value of $t_{0}/a^2$ for the ensemble \texttt{C420a00b334}.}
\label{fig:Correction_t0}
\end{figure}

The result of the isospin breaking correction to $t_{0}/a^2$, for the ensemble \texttt{C420a00b334}, is shown in Figure \ref{fig:Correction_t0}. The distinction between the mass diagram and the sum of the Electromagnetic diagrams shows the actual cancellation both in the signal and in the error with a final result which can be resolved but which is compatible with zero inside of $2\sigma$.

\section{Conclusions \& Outlook}
In this Proceedings, we have presented the results on the continuum and infinite volume scaling of the error of the sea quark IB diagrams for $O\left(a\right)$-improved Wilson fermions and C-periodic boundary conditions in space. We have seen that the gauge error can be reached very easily for the dominant diagrams, i.e. the mass and the tadpole. The statistical error of the lightbulbs becomes relatively small with a moderate number of stochastic sources. Due to SU(3) flavour symmetry, when summed over all flavours, the lantern diagrams vanish identically. By looking at each flavour individually, we have seen that the associated statistical error also becomes quickly sub-dominant compared to the other diagrams.

The volume dependence of the gauge noise follows very well by the asymptotic $\sqrt{V}$ behaviour in the considered range of volumes, i.e. $1.6\ \text{fm} <L<3.2\ \text{fm}$. This means that going to larger volumes, which are relevant for physics, by means of brute-force approaches may prove to be prohibitive. As many before us have noticed, the divergence of the error with the volume can be mitigated by using position-space methods (e.g. by integrating the sea-sea insertions only in subvolumes) and we plan to investigate this idea in the future. On the other hand, the gauge noise does not show the leading divergence in the lattice spacing at least in the considered range, i.e. $0.05\ \text{fm} <a<0.10\ \text{fm}$. Generally, a milder divergence is observed.

We have also observed large correlation between the mass diagram and the electromagnetic ones which resulted in an error reduction for the sum of all contributions for a range of values of the mass shift close to the tuned one. We argue that one can increase the correlation of error by choosing a different discretization for the mass-shift operator, which has the same structure as the tadpole operator.

With this preliminary work we have used only limited statistics and we will extended the analysis to full statistics in the near future. We will also investigate the dependence of the error on the pion mass. In this Proceedings we have shown results only for $t_{0}$ and the meson masses (which are necessary to tune the bare quark mass shift). The RC$^{\star}$ collaboration is also investigating IBE corrections to other observables, e.g. the HVP contribution to the anomalous magnetic moment of the muon (see \cite{Parato:2024} for an update).

\acknowledgments
We acknowledge all the members of the RC$^{\star}$ collaboration for the fruitful discussions and feedback on this work. AC's research is funded by the Deutsche Forschungsgemeinschaft (DFG, German Research Foundation) - Projektnummer 417533893/GRK2575 "Rethinking Quantum Field Theory". The authors gratefully acknowledge the computing time granted by the Resource Allocation Board and provided on the supercomputer Lise and Emmy at NHR@ZIB and NHR@Göttingen as part of the NHR infrastructure. The calculations for this research were partly conducted with computing resources under the project bep00116.

\bibliographystyle{JHEP}
\bibliography{inspire}

\end{document}